\documentstyle[12pt,psfig,epsfig,axodraw]{article}
\advance\textheight by \topskip
\begin{document}
\tolerance=100000
\thispagestyle{empty}
\setcounter{page}{0}

\def\cO#1{{\cal{O}}\left(#1\right)}
\newcommand{\be}{\begin{equation}}
\newcommand{\ee}{\end{equation}}
\newcommand{\br}{\begin{eqnarray}}
\newcommand{\er}{\end{eqnarray}}
\newcommand{\ba}{\begin{array}}
\newcommand{\ea}{\end{array}}
\newcommand{\bi}{\begin{itemize}}
\newcommand{\ei}{\end{itemize}}
\newcommand{\bn}{\begin{enumerate}}
\newcommand{\en}{\end{enumerate}}
\newcommand{\bc}{\begin{center}}
\newcommand{\ec}{\end{center}}
\newcommand{\ul}{\underline}
\newcommand{\ol}{\overline}
\newcommand{\ar}{\rightarrow}
\newcommand{\sm}{${\cal {SM}}$}
\newcommand{\as}{\alpha_s}
\newcommand{\aem}{\alpha_{em}}
\newcommand{\ycut}{y_{\mathrm{cut}}}
\newcommand{\susy}{{{SUSY}}}
\newcommand{\Dir}{\kern -6.4pt\Big{/}}
\newcommand{\Dirin}{\kern -10.4pt\Big{/}\kern 4.4pt}
\newcommand{\DDir}{\kern -10.6pt\Big{/}}
\newcommand{\DGir}{\kern -6.0pt\Big{/}}
\def\Ecm{\ifmmode{E_{\mathrm{cm}}}\else{$E_{\mathrm{cm}}$}\fi}
\def\gluino{\ifmmode{\mathaccent"7E g}\else{$\mathaccent"7E g$}\fi}
\def\photino{\ifmmode{\mathaccent"7E \gamma}\else{$\mathaccent"7E \gamma$}\fi}
\def\mgluino{\ifmmode{m_{\mathaccent"7E g}}
             \else{$m_{\mathaccent"7E g}$}\fi}
\def\taugluino{\ifmmode{\tau_{\mathaccent"7E g}}
             \else{$\tau_{\mathaccent"7E g}$}\fi}
\def\mphotino{\ifmmode{m_{\mathaccent"7E \gamma}}
             \else{$m_{\mathaccent"7E \gamma}$}\fi}
\def\ML{\ifmmode{{\mathaccent"7E M}_L}
             \else{${\mathaccent"7E M}_L$}\fi}
\def\MR{\ifmmode{{\mathaccent"7E M}_R}
             \else{${\mathaccent"7E M}_R$}\fi}
\def\lsim{\buildrel{\scriptscriptstyle <}\over{\scriptscriptstyle\sim}}
\def\gsim{\buildrel{\scriptscriptstyle >}\over{\scriptscriptstyle\sim}}
\def\jp #1 #2 #3 {{J.~Phys.} {#1} (#2) #3}
\def\pl #1 #2 #3 {{Phys.~Lett.} {#1} (#2) #3}
\def\np #1 #2 #3 {{Nucl.~Phys.} {#1} (#2) #3}
\def\zp #1 #2 #3 {{Z.~Phys.} {#1} (#2) #3}
\def\pr #1 #2 #3 {{Phys.~Rev.} {#1} (#2) #3}
\def\prep #1 #2 #3 {{Phys.~Rep.} {#1} (#2) #3}
\def\prl #1 #2 #3 {{Phys.~Rev.~Lett.} {#1} (#2) #3}
\def\mpl #1 #2 #3 {{Mod.~Phys.~Lett.} {#1} (#2) #3}
\def\rmp #1 #2 #3 {{Rev. Mod. Phys.} {#1} (#2) #3}
\def\sjnp #1 #2 #3 {{Sov. J. Nucl. Phys.} {#1} (#2) #3}
\def\cpc #1 #2 #3 {{Comp. Phys. Comm.} {#1} (#2) #3}
\def\xx #1 #2 #3 {{#1}, (#2) #3}
\def\NP(#1,#2,#3){Nucl.\ Phys.\ \issue(#1,#2,#3)}
\def\PL(#1,#2,#3){Phys.\ Lett.\ \issue(#1,#2,#3)}
\def\PRD(#1,#2,#3){Phys.\ Rev.\ D \issue(#1,#2,#3)}
\def\preprint{{preprint}}
\def\Ord{\lower .7ex\hbox{$\;\stackrel{\textstyle <}{\sim}\;$}}
\def\OOrd{\lower .7ex\hbox{$\;\stackrel{\textstyle >}{\sim}\;$}}
\def\MCH {$\tilde\chi_1^+$}
\def \CH{{\tilde\chi}^{\pm}}
\def \LSP{\tilde\chi_1^0}
\def \SNU{\tilde{\nu}}
\def \BARSNU{\tilde{\bar{\nu}}}
\def \MLSP{m_{{\tilde\chi_1}^0}}
\def \MCH{m_{{\tilde\chi}^{\pm}}}
\def \MCHMIN {\MCH^{min}}
\def \ET{\not\!\!{E_T}}
\def \LL{\tilde{l}_L}
\def \LR{\tilde{l}_R}
\def \MLL{m_{\tilde{l}_L}}
\def \MLR{m_{\tilde{l}_R}}
\def \MSNU{m_{\tilde{\nu}}}
\def \PROCESS{e^+e^- \rightarrow \tilde{\chi}^+ \tilde{\chi}^- \gamma}
\def \PI{{\pi^{\pm}}}
\def \DM{{\Delta{m}}}
\newcommand{\bQ}{\overline{Q}}
\newcommand{\ad}{\dot{\alpha }}
\newcommand{\bd}{\dot{\beta }}
\newcommand{\dd}{\dot{\delta }}
\def \CH{{\tilde\chi}^{\pm}}
\def \MCH{m_{{\tilde\chi}_1^{\pm}}}
\def \LSP{\tilde\chi_1^0}
\def \MUL{m_{\tilde{u}_L}}
\def \MUR{m_{\tilde{u}_R}}
\def \MDL{m_{\tilde{d}_L}}
\def \MDR{m_{\tilde{d}_R}}
\def \MSNU{m_{\tilde{\nu}}}
\def \MLL{m_{\tilde{l}_L}}
\def \MLR{m_{\tilde{l}_R}}
\def \mhf{m_{1/2}}
\def \MST{m_{\tilde t_1}}
\def\tth{\tilde{t}\tilde{t}h}
\def\qqh{\tilde{q}_i \tilde{q}_i h}
\def\t1{\tilde{t_1}}
\def \pt{p{\!\!\!/}_T}  
\def\lapp{\mathrel{\rlap{\raise.5ex\hbox{$<$}}
                    {\lower.5ex\hbox{$\sim$}}}}
\def\gapp{\mathrel{\rlap{\raise.5ex\hbox{$>$}}
                    {\lower.5ex\hbox{$\sim$}}}}
\vspace*{\fill}
\vspace{-0.5in}
\begin{flushright}
{December,2001}\\
{hep-ph/0112182}
\end{flushright}
\begin{center}
{\Large \bf
Four Body Decay of the  Stop Squark
at the Upgraded Tevatron}\\[1.00
cm]
{\large Siba Prasad Das$^{a,}$,\footnote{\it spdas@juphys.ernet.in} 
Amitava Datta$^{a,}$} \footnote{\it adatta@juphys.ernet.in}\\[0.3 cm]
{and}\\[0.3 cm]
{\large Monoranjan Guchait$^{b,}$} \footnote{\it monoranjan.guchait@cern.ch;
Present address: Department of High Energy Physics, Tata Institute of 
Fundamental Research, Homi Bhabha Road, Bombay-400005, India.}
{\\[0.3 cm]}
{\it $^a$ Department of Physics, Jadavpur University, 
Calcutta-700032,India}\\[0.3cm]
{\it $^b$ The Abdus Salam International Centre for Theoretical Physics,
Strada Costieara 11, I-34014, Trieste, Italy}\\[0.3cm] 

\end{center}
\vspace{.2cm}

\begin{abstract}
{\noindent\normalsize 
We investigate the prospect of stop squark search at Tevatron RUN-II
in mSUGRA motivated as well as unconstrained supersymmetric models,
when the lighter stop squark turns out to be the next lightest 
supersymmetric particle (NLSP). In this case the decay into a 4-body 
final state consisting of a b quark, the lightest neutralino and two 
light fermions may compete with the much publicized loop induced two 
body decay into a charm quark and the lightest neutralino. We 
systematically study the parameter space in mSUGRA where the
lighter stop squark turns out to be the NLSP and calculate the 
branching ratios of the competing channels in both models. Our
results show that the four body decay may indeed be the main discovery
channel particularly in the low tan $\beta$ scenarios. We discuss the 
detectability of stop squark pairs in the 4-body decay channel
leading to one lepton with 2 or more jets accompanied by a large amount 
of missing energy. We also studied the corresponding background processes 
and the kinematic cuts required to suppress them using parton level Monte 
Carlo simulations. We have commented upon with illustrative examples, 
the required revision of the existing mass limits of the stop NLSP assumed 
to decay solely into the loop induced 2-body channel in the presence of 
the competing 4-body decay.
}
\end{abstract}
\vskip1.0cm
\noindent
\vspace*{\fill}
\newpage

\section*{I.~Introduction}
\label{sec_intro}
The Minimal Supersymmetric Standard Model (MSSM) ~\cite{susy} is a  well 
motivated extension of the Standard Model(SM). As of now, neither there is 
any evidence of it nor has it been ruled out by the
electroweak precision 
measurements at LEP ~\cite{leppre}. Unfortunately, we are not equipped with
any information  about the range of superparticle masses from 
the theoretical 
point of view. On the other hand, from the experimental side, there are some
lower bounds from the non observation of superparticles in colliders,
 such as LEP~\cite{lepbound} and Tevatron RUN-I~\cite{xsusy}. 

The up type squark of the  third generation - the stop squark, the 
superpartner
of the top quark - is something special. It is because, of the large top 
Yukawa 
coupling which controls the evolution of the soft-supersymmetry breaking
masses of the left and right handed stop squarks,
 $\tilde t_L, \tilde t_R$, via the  Renormalisation Group equations.
This tends to reduce these masses , compared   
to  the other squark masses~\cite{sugra}. Moreover, because of the large 
top quark
mass, the two weak states $\tilde t_L, \tilde t_R$ may mix very strongly 
leading to a  relatively large splitting between the two physical mass 
eigenstates 
$\tilde t_1, \tilde t_2~$~\cite{stopmix}
(in our notation $m_{\tilde t_2} >
m_{\tilde t_1}$). 
Interestingly, the mass of the lighter states $\tilde t_1$ may be even below 
the top mass. In fact, it is quite conceivable that it happens to
be the next lightest supersymmetric particle (NLSP), the lightest 
neutralino $\tilde\chi_1^0$  being  the lightest supersymmetric 
particle (LSP) by assumption.

In the coming years the second phase of Tevatron experiments,
the RUN-II, will start operating with an integrated 
luminosity of at least
2 fb$^{-1}$ per experiment at 2 TeV center of mass energy, which is about 
ten times larger than the acquired luminosity in RUN-I with center of 
mass energy 1.8 TeV. With a further luminosity upgrade
it is expected to collect a luminosity of something
like 15-20~fb$^{-1}$ after a few years of operation. 
The phenomenology of the stop squark could be  of
special interest, since it might be the only strongly interacting
 sparticle within the kinematic range of RUN-II experiments. It
is, therefore, very important to fix up the strategies for isolating the stop 
signal for all conceivable decay modes.

Currently, the search for stop at LEP~\cite{lepbound} 
and Tevatron RUN-I~\cite{tevbound} experiments have yielded  negative 
results. 
The most stringent
bound comes from Tevatron experiments which puts a lower 
limit on lighter stop mass $m_{\tilde t_1} \ge $ 119 GeV for 
$m_{\tilde\chi^0_1}=$40 GeV 
and this limit becomes slightly weaker 
for higher value of $m_{\tilde\chi^0_1}$, e.g, $m_{\tilde t_1} \ge $ 
102 GeV for
$m_{\tilde\chi^0_1}=$50 GeV~\cite{tevbound}. In deriving these 
limits, it was assumed that the loop
induced, flavor changing decay into a charm quark and the
LSP~\cite{hikasa}, 
\begin{equation} 
\tilde t_1 \ar c \tilde\chi_1^0
\label{loopdk}
\end{equation}
occurs with 100\% branching ratio(BR) - until recently assumed 
to be a valid assumption if the  
$\tilde t_1$ state happens to be the NLSP. 
In the R parity conserving model, LSP cannot 
decay further and escapes the  detector resulting a large imbalance 
of transverse energy. The stop signal is tagged through identifying 
jets and missing energy. Since the production of stop pairs is dominantly 
via QCD and depends on its mass only, the above limits from the Tevatron 
are fairly model independent, except for the dependence on 
$m_{\tilde\chi_1^0}$, which influences the efficiency of the kinematical cuts.

In the minimal supersymmetric extension of the standard model with
arbitrary soft breaking  masses of the superparticles, the stop may  
be the NLSP.
However, the stop NLSP may also be realized in more constrained models
like the minimal version of the supergravity model (mSUGRA). The latter is the 
most economical model in the context of supersymmetry (SUSY) searches
in colliders. 

In the mSUGRA model, the 
supersymmetry breaking takes place in hidden sector which is communicated
to the visible sector by gravitational interactions. The mass spectrum of 
sparticles at a lower energy scale can be obtained by the  
Renormalisation Group equations from the inputs at some higher 
scale, which is usually assumed to be the Grand Unification theory(GUT) scale
($M_G$). In mSUGRA models, these input at $M_G$
are the common scalar mass ($m_0$), the common gaugino mass
($m_{1/2}$), the tri-linear scalar coupling term $(A_0)$, 
$\tan\beta$ (the ratio of  two vacuum expectation values of the two higgs doublets which generate the masses
of the fermions and gauge bosons through electroweak symmetry breaking) and the sign of $\mu$, 
(the higgsino mass parameter)~\cite{sugra}.
The magnitude of $\mu$ is fixed by the
radiative electroweak symmetry breaking condition (REWSB)~\cite{rewsb}. 
Surprisingly due to  the interplay of 
these parameters, which we will discuss later, there is a substantial
region of mSUGRA parameter space where the $\tilde t_1$ happens to be the  
NLSP and sometimes the only squark within the striking range of the 
Tevatron.

 The authors of
~\cite{boehm,djouadi} have emphasized that a competiting channel of 
$\tilde t_1$ decay may  
be there even if the stop is the NLSP. 
The 4-body decay of $\tilde t_1$ into a b quark, the LSP and 
two approximately massless fermions~\cite{boehm,djouadi}, 
\be 
\tilde t_1 \ar b \tilde\chi_1^0 f \bar f' 
\label{4dk}
\ee
via heavier SUSY particles, may also  
have significant decay rates, or even dominate over the loop
induced decay mode, eq.\ref{loopdk}, for certain
regions of the  MSSM or mSUGRA parameter space. As a consequence, this
4-body decay mode along with stop pair production in colliders 
may yield final states containing a
lepton plus jets accompanied by huge amount of missing energy 
\footnote {It is to be 
noted that similar final states occur from the production of other SUSY 
particle ( e.g., squarks, gluinos) and their subsequent cascade 
decays in colliders~\cite{xsusy}. These  signals are  used in conventional 
SUSY searches.}. This event topology appears to be indeed
different in comparison to the final state  
 resulting from the loop induced stop decay eq.~\ref{loopdk}.
The new observation on the 4-body decay, therefore, necessitates
revision  of the 
strategy to detect the signal of $\tilde t_1$ when it is the NLSP.

In the framework of most of the SUSY models, stop squarks has many
interesting decay modes depending on its mass. If it is sufficiently
heavy, 
then the main decay modes occurs through top quarks and neutralinos,
\be
\tilde t_1 \ar t \tilde\chi_j^0~(j=1-4) 
\label{2topdk}
\ee
if it is kinematically allowed. 
There is also another competitive decay channel
through charged current interactions into bottom quarks and a lighter
chargino($\tilde\chi_1^\pm$),
\be
\tilde t_1 \ar b \tilde\chi_1^+  
\label{2dk}
\ee
If these modes are not kinematically accessible, but still 
$m_{\tilde t_1}$ heavy enough, then the  
3-body decay modes to bottom quarks, a $W$ boson or 
a charged higgs scalar $H^\pm$ and neutralinos,
$\tilde t_1 \ar b W^+ \tilde\chi_j^0$ 
or $\tilde t_1 \ar b H^+ \tilde\chi_j^0$ (j=1-4), can be 
accessible~\cite{porod1}. 
These decay channels, particularly, the $W$ boson final states may 
be dominant in absence
of the 2-body decay channels, eq.(\ref{2topdk},\ref{2dk}). 
Beside these 3-body
decay modes, in the light slepton scenario which may arise in some 
mSUGRA models, lighter stop can decay with 
a final states containing 
sleptons~\cite{hikasa,boehm,guchait,porod2},
\be
\tilde t_1 \ar b \ell \tilde\nu ~;~ \tilde t_1 \ar b \tilde\ell \nu
\label{3dk}
\ee

The search prospects of 
$\tilde t_1$ state at Tevatron experiments has been investigated
by many authors~\cite{xsusy,guchait,tata,lykken,oliver,yaan}.  
These studies
were carried out by considering the 2-body decay mode of $\tilde t_1$
into a b quark and a lighter chargino, eq.\ref{2dk},
yielding a single lepton or a dilepton pair plus large amount of 
missing energy with some
hadronic activities following the cascade decays of $\tilde\chi_1^+$
into a LSP and massless fermions, 
$\tilde\chi_1^+ \ar \chi_1^0 f \bar f'$. Though this final state has
the same particle content as in eq.2, the kinematical characteristics
are quite different in the two cases. Hence a slightly different set of 
kinematical cuts is needed for isolating the channel, in eq.2, 
from the background.

Similar studies were also
carried out for the same signal with different kinematics in the light 
slepton scenario
where $\tilde t_1$ decays via 3-body into a b quark and a charged 
slepton (lepton) and neutrino (sneutrino), eq.\ref{3dk}.
The SUSY searches take a dramatic turn in the high $\tan\beta$ regime
where lighter staus($\tilde\tau_1$), the SUSY partner of tau lepton
turns out to be very light, even may be the NLSP. It yields a huge
number of tau leptons in the SUSY particle production and their
cascade decays in colliders~\cite{baer}. In  Ref.~\cite{yaan}
discovery potential of $\tilde t_1$ states has been studied for Tevatron 
RUN-II in high $\tan\beta$ regime.

However, the discovery potential of $\tilde t_1$ of the stop  NLSP decaying 
dominantly into the 4-body final state, especially in the context of the 
mSUGRA model, has not yet been studied systematically for the upgraded 
Tevatron. The purpose of this work is to investigate this possibility 
further. In our study, we concentrated on the channel with  a single 
lepton plus 2 or more jets accompanied by large amount of missing energy. 
This final state appears from stop pair production following the semileptonic
decay of one $\tilde t_1$ and hadronic decay of the other. We also discuss 
the possible sources of SM backgrounds and optimize the set of cuts to 
minimize the contamination. 

Our investigation is based on models which conserves R parity. In  
R parity breaking models, few more channels of $\tilde t_1$ decay open up 
which depend on R parity violating couplings. The collider 
phenomenology of $\tilde t_1$ in presence of R parity breaking SUSY 
model will be discussed elsewhere~\cite{spdas}.

We have organized our paper as follows. In Sec. II, we have isolated 
the region of the parameter space where lighter stop happens to be the NLSP
in the mSUGRA model. In Sec. III, the key points of stop decay 
patterns have been mentioned. The parameter spaces, both in mSUGRA
and MSSM, where the branching ratio of the 4-body decay is
significant, are identified. In Sec. IV, we investigate the signal 
of $\tilde t_1$ in the 4-body decay channel by triggering single lepton 
plus jet with missing energy channel. We also estimate the 
corresponding SM backgrounds after minimizing them by imposing kinematical 
cuts. We conclude in Sec.V with a few remarks. We have also  
commented on the possibility of revision in the current mass 
limits in the presence of the competeting 4-body decay mode along with
illustrative examples.

\section*{II.~Stop NLSP in the mSUGRA model}
\label{sec_msugra}

In this section we focus our attention on the parameter space in
the mSUGRA model where the stop is the NLSP. As discussed in the introduction
the model is chracterised by five free
parameters: $m_0$, $m_{1/2}$, $A_0$, tan$\beta$ and sign$(\mu)$.

For a given $m_0$, $m_{1/2}$, tan$\beta$ and sign$(\mu)$, the stop
turns out to be  the NLSP for a range of $A_0$
($|A_{0min}|\le|A_0|\le |A_{0max}|$). The sign of  $A_0$ is required 
to be the opposite to that of $\mu$ in order to produce larger mixing 
in the stop mass matrix. For $|A_0|\le |A_{0min}|$ the mixing is small 
and the lighter stop becomes heavier than the chargino. For $|A_0|\ge 
|A_{0max}|$, the lighter stop becomes lighter than the lightest neutralino 
which is also the lightest supersymmetric particle. An additional 
constraint on $|A_{0max}|$ exist in principle due to the requirement that the
vacuum must not break charge and color symmetry ( the CCB condition)
~\cite{ccb}.
However, quite often this constraint becomes redundant in the 
mSUGRA model since the requirement that the stop be heavier 
than the LSP produces 
a stronger bound at least in the low $m_0$, $m_{1/2}$ scenario.
Moreover the upper bound is in general unimportant for this paper, 
since the BR of the 4-body decay is
significant only when $\MST$ is close to $m_{{\tilde\chi}_1^\pm}$ 
so that the chargino is of relatively small virtuality. 
This happens for $A_{0}$ close to $A_{0min}$.

\begin{figure}[!t]
\vspace*{-3.5cm}
\hspace*{-3.0cm}
\mbox{\psfig{file=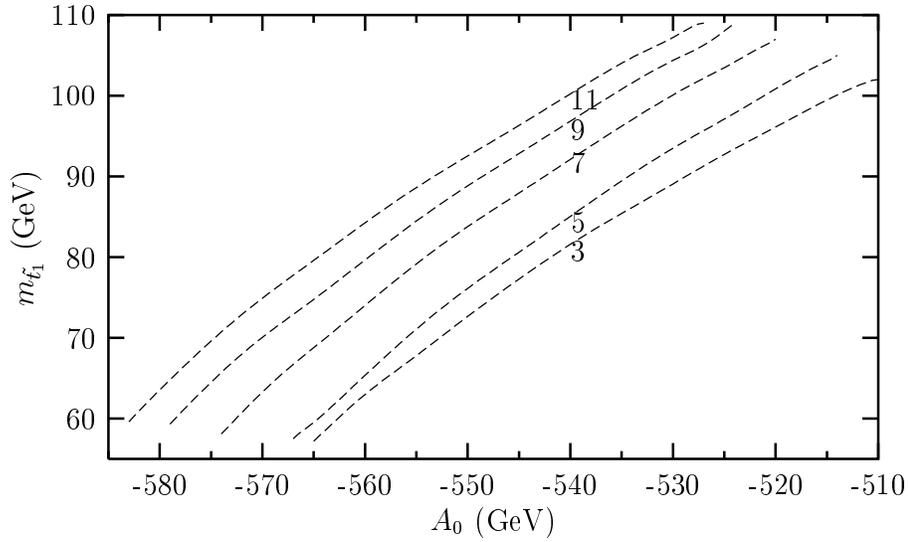,width=20cm}}
\vspace*{-16.7cm}
\caption{\small 
Trilinear coupling ($A_0$) vs. Masses of the 
stop NLSP ($\MST$) for 
tan$\beta$=3,5,7,9,11
where $m_0$=200, 
$\mhf$=145, 
and sign($\mu$)=+ve 
in mSUGRA model.
}
\label{fig_1aovsmt}
\end{figure}

\begin{figure}[!t]
\vspace*{-3.5cm}
\hspace*{-3.0cm}
\mbox{\psfig{file=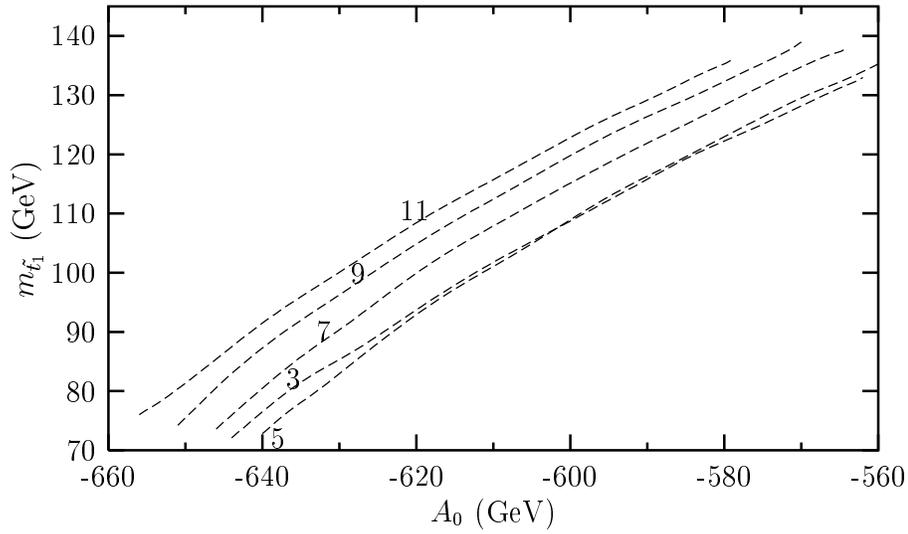,width=20cm}}
\vspace*{-16.7cm}
\caption{\small 
Same as in Fig.1, but for  
$m_0$=140,
$\mhf$=180, and sign($\mu$)=+ve. 
}
\label{fig_2aovsmt}
\end{figure}

For illustration, in Fig.1 we present the stop mass  
as a function of $A_0$ for different tan$\beta$, 
$m_0$=200, $m_{1/2}$=145, and $\mu>0$ ({\bf  all masses and
mass parameters in this paper are in GeV}). Here
$m_{{\tilde\chi}_1^\pm}$=106
for tan$\beta$=5 which varies modestly with tan$\beta$.
Only those values of A  are considered for which the 
$\tilde t_1$ is the NLSP.
As seen from the figure  $|A_{0max}|$ nearly equal to 565 and
$|A_{0min}|$ nearly equal to 510 for tan$\beta$=3.
For a given $A_0$, the  mass of the lighter stop is larger for
larger
tan$\beta$ since mixing is reduced by the  $\frac{\mu}{\tan\beta}$ 
in the stop mixing matrix.

Similar information for
$m_0$=$140$, $m_{1/2}$=180, $\mu>0$ is given in Fig.2. In Fig.2
we find that higher masses of the stop NLSP are  allowed since here
$m_{{\tilde\chi_1}{\pm}}$ is somewhat larger. Here $\MCH$=136
for tan$\beta$=5 which varies modestly with tan$\beta$.

If $m_0 >> \mhf$, then $m_{\tilde t_1}$ tends to increase, unless $A_0$
is properly tuned so that a large mixing in the stop mass matrix
still leads to a low mass stop NLSP. Beyond some large value of
$m_0$ extreme fine tuning is needed which is rather unaesthetic.

On the otherhand for $\mhf >> m_0$, quite often a large $A_0$
is needed to produce a stop NLSP, which is in conflict with
the invariance of the vacuum under charge-color symmetry ~\cite{ccb}. 
Even if the CCB condition is disregarded, we find that in a
very narrow 
range of $A_0$ we have the stop NLSP. Otherwise a slepton
NLSP scenario is obtained over most of the parameter space.
Thus in this paper we shall focus our attention on scenarios
with $m_0 \sim \mhf$.

It is clear from Fig.1 and Fig.2 that if the BR of the loop induced 
decay is indeed $\sim$ 100\% as is normally assumed, then the limits from LEP and Tevatron 
are already sufficiently restrictive and can exclude a large region 
of the mSUGRA parameter space ( in particular certain
ranges of $A_0$ for 
given values of other mSUGRA parameters will be excluded ). It is also 
reasonable to hope that RUN-II at a much higher luminosity will extend 
this probe to a much larger region of the parameter space. However,
as emphasized in the introduction, the 4-body decay, eq.2, can indeed 
compete with the loop induced 2-body decay and reduce its
BR significantly ~\cite{boehmi,djouadi}. This will be illustrated with several 
examples in the context of 
the mSUGRA model in the next section. The existing limits on 
$m_{\tilde t_1}$, therefore, may require a revision. 
{\bf We shall try to illustrate estimate the 
required revision in section III}. More importantly, when new and stronger 
limits come from RUN-II, the interplay of the two competing channels should be
kept in mind. Such limits will depend not only on $\MST$ and
$\MLSP$, but also on the relative BRs of the two competing
channels for stop decay and, hence, on other SUSY parameters.

\section*{III.~Stop decay branching ratios}
\label{stopdecay}
As mentioned in Sec.I,
when the $\tilde t_1$ is the NLSP,
two decay channels are allowed  which are  competitive with
each other - the loop induced flavor changing decay mode,
eq.\ref{loopdk}~\cite{hikasa} and 4-body final 
states with nearly  massless fermions, the bottom quark and the LSP,
eq.\ref{4dk}~\cite{boehm,djouadi}.

The 4-body decay mode of lighter stop, eq.\ref{4dk}
occurs through many diagrams mediated by: (a) $W$ and/or
 $H^\pm$ with virtual
$t$, $\tilde b$ and $\tilde\chi^\pm_{1,2}$, (b) virtual 
$\tilde\chi^\pm_{1,2}$ with $\tilde\ell$ or $\tilde\nu_\ell$. 
The dependence of 4-body decay rates on supersymmetric parameters
has been discussed in great detail in the paper of  
Ref.~\cite{boehm, djouadi}. Nevertheless, few comments are in order:

\begin{itemize}
\item
Among all the diagrams, only few have significant 
contributions. As for example, the diagrams mediated by $H^\pm$ and 
$W$ are heavily suppressed because, $m_{H^\pm}$ are larger than
$M_W$ and Yukawa coupling between $H^\pm$ and fermions are suppressed 
by respective fermion  masses. In addition to these, the diagrams mediated by
squarks have very little contribution as those are expected to be
larger($\ge$250) ~\cite{pdg} than $\tilde t_1$ state.   
The top quark mediated diagrams do not give large contributions unless
$m_{\tilde t_1} \sim m_t + m_{\tilde\chi_1^0}$.           

\item
The diagrams which expected to give a significant contribution  
to the 4-body decay rate are mediated by charginos 
$\tilde\chi_i^\pm$(i=1,2) and sleptons ($\tilde\ell,\tilde\nu_\ell$).
When the virtuality between $\tilde t_1$ and $\tilde\chi_1^\pm$,
$\tilde\ell$ or $\tilde\nu_\ell$ is not very large then these 
diagrams do give a large contributions. As for example, the diagrams
involving tau sleptons which could be  rather light because of mixing
(especially, for high $\tan\beta$~\cite{yaan}) may  contribute appreciably.
\end{itemize}

In the MSSM the soft breaking parameters can be chosen arbitrarily.
One can, therefore,  suppress the loop decay to the extent one requires.
Now if the stop is chosen to be the NLSP, the 4-body decay will 
automatically have a large branching ratio. For the detectable signal,
which we consider, one also needs a significant leptonic BR of the
stop (see section IV ). This can be achieved by choosing 
relatively light sleptons. In the following we shall illustrate this 
with different inputs.

What, however, is interesting is that even in a constrained model like
mSUGRA, one can have a significant BR for the 4-body decay in
some  appropriate region of the parameter space. Moreover, a significant
leptonic BR can be accommodated in a small but interesting 
region.

We first consider the unconstrained MSSM model.
As is well-known the loop decay width is controlled by the
parameter $\epsilon$ which denotes the amount of
$\tilde{t}_{L,R}$--$\tilde{c}_L$ mixing ~\cite{hikasa} and enters 
in the decay width formula as,
\be
\Gamma(\tilde t_1 \ar c {\tilde\chi_1^0}) = \frac{\alpha}{4}
|\epsilon|^2 f^2 m_{\tilde t_1} \left(1 - \frac{m_{\tilde\chi_1^0}^2}{
m_{\tilde t_1}^2}\right)^2
\ee
The detailed expressions for $\epsilon$ and the function {\it f} can be found 
in Ref.~\cite{hikasa,boehm} and Ref.~\cite{boehm,guchait} respectively. 
If the mixing angle $\theta_t$ in the stop
sector is chosen appropriately so that it yields very small value of the 
$\epsilon$ parameter, the loop decay is reduced 
drastically~\cite{boehm,djouadi}. This effect is shown in Fig.3. The choice
of the SUSY parameters for this figure are as follows assuming gaugino mass
unification:  
the  SU(2) gaugino mass parameter $M_2$=180,
the higgsino mass parameter $\mu$ = +400,
tan$\beta$=4. This three parameters completely defines 
the Chargino and Neutralino sector. 
We have taken CP-odd neutral higgs mass $M_A$=150 which requires to calculate
the neutral higgs to obtain the value of the $\epsilon$ parameter. 
The  others MSSM parameters requires for calculating the stop 
branching ratios are the following: the common
scalar squark mass $m_{\tilde q}$=500, common slepton mass
$m_{\tilde\ell}$=175, the trilinear coupling in the bottom sector
$A_{b}$=300, the trilinear coupling in the tau sector 
$A_{\tau }$=200. With these parameters, $\cos{\theta_{\tilde t}}$=0.11
which leads to a very small $\epsilon$.
Because of this
the loop decay BR sharply falls as soon the 4-body decay is 
kinematically allowed. The total BR for the four body decay 
exceeds the loop decay BR for $\MST$ $\approx$ 95 and approaches 
100\% for slightly higher values of $\MST$. For our parameter
set $\MCH \approx $ 162. Thus the 4-body BR can be rather 
large even for charginos of relatively high virtuality. We have 
checked that even if this smaller value of the angle $\theta_{\tilde t}$ 
is not used so that
$\epsilon$ is somewhat larger,
the enhancement of the 4-body decay BR, though less dramatic,
still exist over a significant region of the parameter space. Moreover 
the leptonic BR of the stop is greater than 10\%. This, as
we shall see in the next section, as adequate for the detection
of the 4-body decay of the stop in the channel studied by us.

We are now in a position to rexamine the possible impact of 
the large BR of the 4-body decay channel on the current and future 
limits on $m_{\tilde t_1}$. For the choice of parameters in Fig.3, 
$\MLSP$=86.41. For such a large value of the LSP mass, no limit 
on $\MST$ is currently available. However, 
the expected  mass limits from at RUN-II needs careful handling. As we 
see from Fig.3 the total BR of the 4-body decay approaches 100\%
for $\MST \ge$ 100. Obviously, the searches for $\tilde t_1$ NLSP in this 
mass range should be based on the 4-body decay channel.

\begin{figure}[!t]
\vspace*{-3.5cm}
\hspace*{-3.0cm}
\mbox{\psfig{file=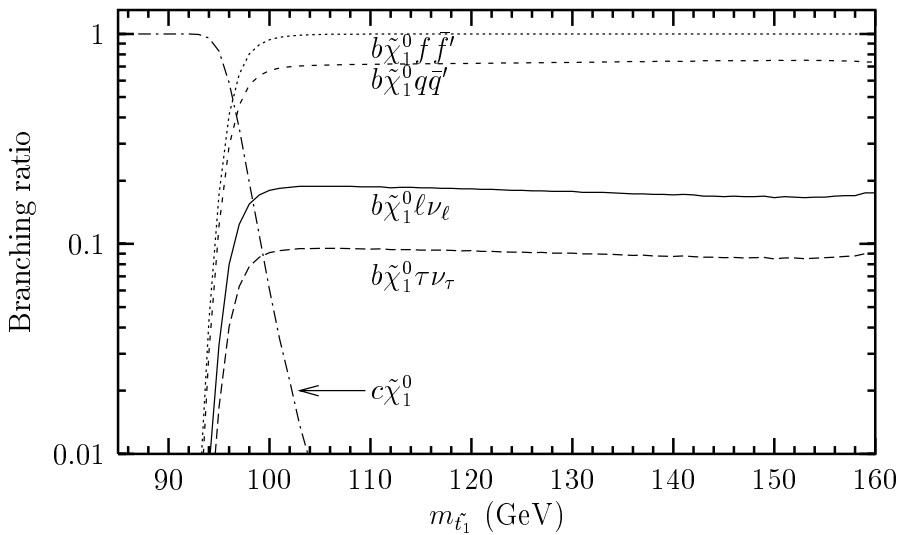,width=20cm}}
\vspace*{-16.7cm}
\caption{\small The branching ratio BR($\tilde t_1 \ar b \tilde\chi_1^0 f 
\bar f'$) 
and BR($\tilde t_1 \ar c \tilde\chi_1^0$) as a function of 
the stop squark mass ($\MST$) in MSSM scenario, 
where the common
scalar squark mass $m_{\tilde q}$=500, common slepton mass
$m_{\tilde\ell}$=175, $M_2$=180, $\mu$ = 400,
$A_b$=300, $A_{\tau}$=200, $M_A$=150.
$\tan\beta$=4, and $cos_{\theta_t}$=0.11.
}
\label{fig_mssm1}
\end{figure}
\begin{figure}[!t]
\vspace*{-3.5cm}
\hspace*{-3.0cm}
\mbox{\psfig{file=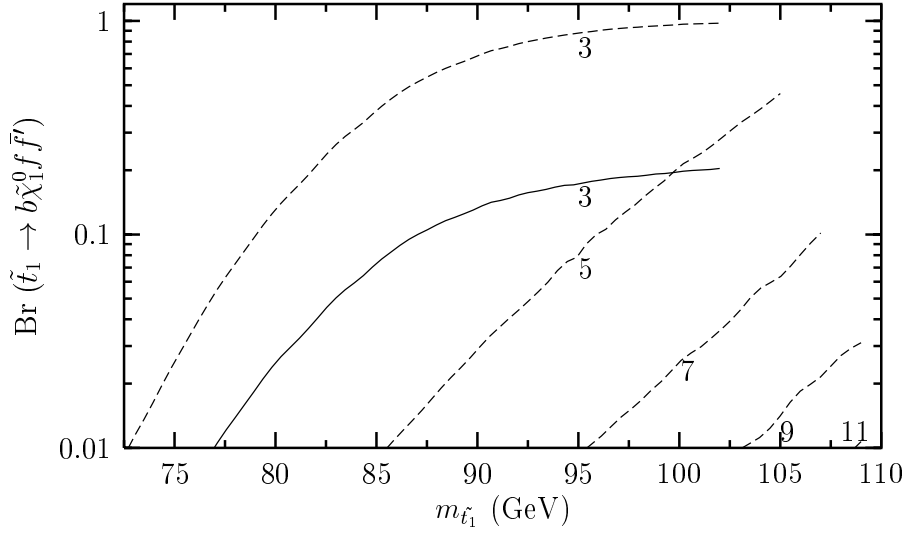,width=20cm}}
\vspace*{-16.7cm}
\caption{\small 
The branching ratio BR($\t1 \ar b \tilde\chi_1^0 f \bar f'$)
as a function of stop ($\MST$) for tan$\beta$=3,5,7,9,11 (dashed
lines) in mSUGRA model, where $m_0$=200,
$\mhf$=145, and sign($\mu$)=+ve. The solid line is the 
leptonic branching ratio for tan$\beta$=3.
}
\label{fig_5mtbr}
\end{figure}
\begin{figure}[!t]
\vspace*{-3.5cm}
\hspace*{-3.0cm}
\mbox{\psfig{file=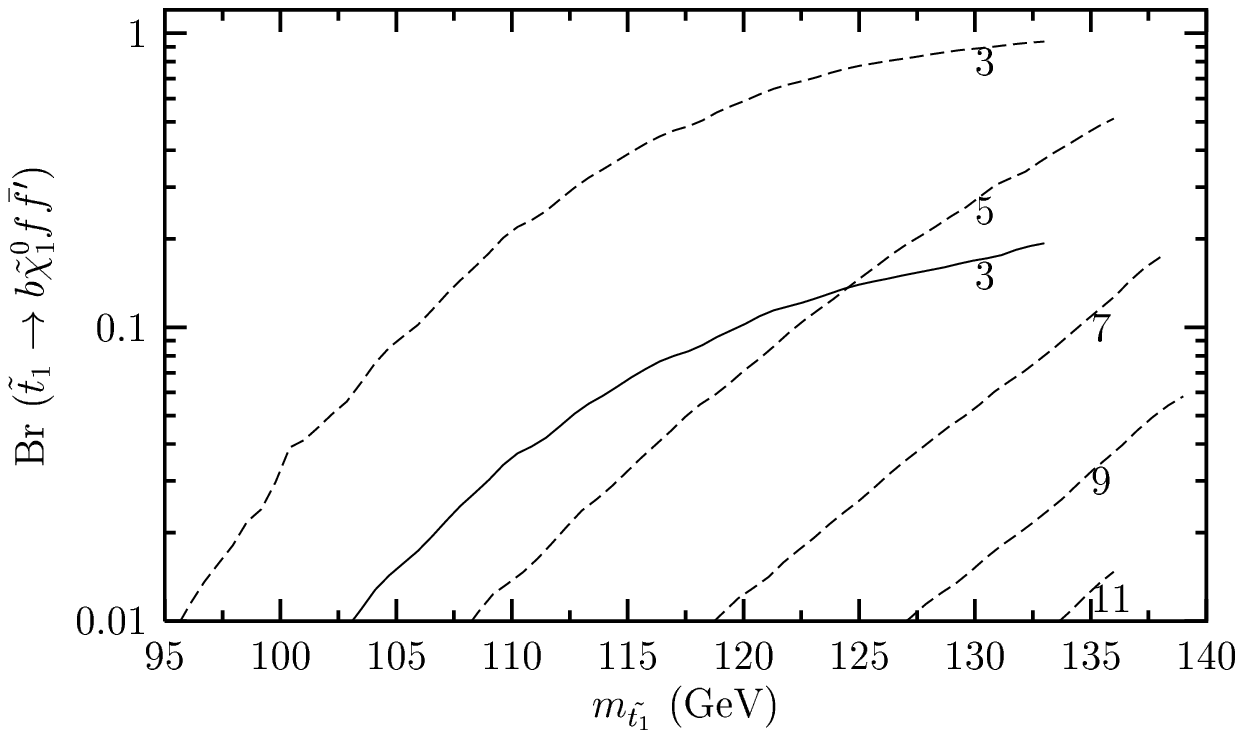,width=20cm}}
\vspace*{-16.7cm}
\caption{\small 
Same as in Fig.4, but for $m_0$=140, $\mhf$=180 and sign($\mu$)=+ve.
}
\label{fig_4mtbr}
\end{figure}

We now turn our attention to the 4-body BR in the mSUGRA model.
We present the 4-body
BR as a function of $\MST$ (Fig. 4 and 5). The choices of SUSY
parameters are as in Fig.1 and Fig.2 respectively. In Fig.4, 
$\MCH \approx$ 103,  although its precise value depends on
tan$\beta$. 
We find that the BR is appreciable when low values of tan$\beta$
are considered and the virtuality of the chargino is small.
This is because
the loop decay rapidly increases with tan$\beta$. Thus total 
4-body BR $>10\%$ is
allowed only for tan$\beta$ $\lsim $ 7. 
Thus the phenomenology of the stop NLSP 
and its 4-body decay is interesting for a rather limited region of 
the $A_0$-tan$\beta$ parameter space.

For the set of parameters in Fig.4, $\MLSP \approx $55. In this
scenario the current mass limit from the Tevatron based on the 
loop decay is $\MST \gsim $100. However, in the tan$\beta$=3 scenario
the total BR of 4-body decay is quite sizable for $\MST \gsim $85.
Hence masses significantly lower than 100 cannot be excluded 
apriorily. It follows from 
Fig.4 that for tan$\beta$ $\lsim$ 5 the 4-body decay may indeed play 
a significant role in the discovery of $\t1$. In fact for the 
entire range 3 $\lsim$ tan$\beta$ $\lsim$ 5 both the decay 
modes eq.(\ref{loopdk},\ref{4dk}) of the stop NLSP should be taken 
into account. Negative 
results from stop search on the other hand, already excludes
certain ranges of the $A_0$ parameter. For example, the low BR 
of the 4-body decay for tan$\beta$ $>$5 indicate that 
the current limits from Tevatron RUN-I based on the loop decay of
$\tilde t_1$ are valid for these tan$\beta$. Referring back to Fig.1, we 
find that, e.g, $A_0 \gsim$525 is excluded for tan$\beta$=7 and 
$m_0$, $\mhf$, sign($\mu$) as given above. This limit is certainly 
stronger than the one obtained from  the CCB condition or from the 
requirement that stop be  heavier than the LSP. Moreover, 
the limit from the CCB condition does not hold if the
SU(2)$\otimes$U(1) breaking minimum happens to be a false 
vacuum with a life time larger than the age of the 
Universe~\cite{fvacuniv}. In 
contrast the limits from collider searches are free from this
ambiguity. The  upper limit on $A_0$ can be further strengthened from
stop searches in the $\tilde t_1 \ar b \tilde\chi_1^+$ channel.

The conclusion from Fig.5 are similar except for the fact that
the allowed values of $\MST$ are somewhat larger in this case.
From Fig.5 it follows that tan$\beta \gsim$ 5 the current limits from 
RUN-I are valid. Combining the results of Fig.2 and Fig.5 we 
find , e.g., $|A_0| \gsim$ 600 for tan$\beta$=7 violates the CDF
contraint on the stop mass.

In order to get a viable signal from  4-body stop decay it is
important that the BR of decays into electrons and muons be
significant. This requirement further restricts the interesting
region of the
parameter space. For a given $\mhf$ this BR decreases as $m_0$ is
increased. This
happens simply because the sleptons get heavier. Of course when
$m_0$ is
sufficiently large the BR becomes independent of $m_0$ since in
this
situation only the W - exchange diagram contributes. In both
Fig.4 and Fig.5, we find that it is possible
to obtain the leptonic BR
$\ge 10\%$ for suitable choices of the SUSY parameters. This may be
adequate for the detectability of the 4-body stop decays as we shall see in
the next section.

\section*{IV.~Stop production: Signal and Backgrounds}
In hadron colliders, stop pairs are produced via gluon-gluon fusion and
quark-antiquark annihilation as,
\be
gg,q \bar q \ar \tilde t_1 \tilde t_1^*
\ee
The production cross section depends only on the mass of 
$\tilde t_1$ without any dependence on the mixing angle in the 
stop sector, since it is a pure QCD process. 
The total pair production cross section 
at the Tevatron for $\sqrt s=$ 2 TeV is $\simeq$ 15-0.3 pb which is 
40\% larger than the cross section for $\sqrt s=$ 1.8 TeV, for the range
of $m_{\tilde t_1}\sim$ 100-200. The QCD 
corrections enhance this cross section by $\sim$30\% over most of 
 SUSY parameter space accessible at 
Tevatron~\cite{spira}.

We investigate the signal of stop pair 
production in the channel  
$\ell+$jets($\ge 2)+p{\!\!\!/}_T$, 
assuming  that one stop decays
leptonically and the other hadronically, i.e ,
\be
\tilde t_1 \ar b \tilde \chi_1^0 \ell \nu_\ell \\ \nonumber 
~ ; ~\tilde t_1^* \ar b \tilde\chi_1^0 q \bar q', 
\label{stopsig}
\ee
where q=u,d,c,s and $\ell=e,\mu$. 
The same event topology also appears from top pair 
production,
\be 
p \bar p \ar t \bar t;~~ t \ar \bar b q \bar q',
~~\bar t \ar  b q \bar q'
\label{ttbg}
\ee 
with a $\frac{4}{27}$ branching ratio suppression. 
 Other main sources of non negligible SM backgrounds
come from $W$ boson pair production accompanied by QCD jets, 
$p\bar p$ $\ar$ W$^+$W$^-$, W+2jets, WWj, with  one W $\ar \ell \nu_\ell$ 
and jets coming either 
from the other W or from QCD shower. 

We have analyzed the signal and 
background cross section using parton level Monte Carlo simulation
without taking into account the fragmentation effects of jets. In our 
calculation we set renormalisation and fragmentation scale, $Q^2=
\hat s$ and used CTEQ4M~\cite{cteq} for the parton distributions 
in proton. The energy of the visible particles  in the signal,
depends on $\Delta m=m_{\tilde t_1} - m_{\tilde\chi_1^0}$;
the larger the value of $\Delta m$, the harder is the momentum of leptons and jets in the final state and one obtains a better efficiency 
of the cuts. We computed the signal cross section for a wide 
range of $m_{\tilde t_1}$ values varying $m_{\tilde\chi_1^0}$ 
by fixing r, $r=\frac{m_{\tilde t_1}}
 {m_{\tilde\chi_1^0}}$,  where r$>1$.

The cross section for the process $p \bar p \ar W +$ 2 jets 
has been estimated using MADGRAPH
~\cite{madgraph}. We have cross checked the cross sections with the 
numbers quoted in Ref.~\cite{vecbos} and found that they are  
consistent within a few percent.
Similarly, we also generated the process  $p \bar p \ar$W~W~j using the 
same code. 
 
For the selection of events, we use the following  
cuts for the transverse momentum $p_T$, the rapidity $\eta$ and missing
energy $p{\!\!\!/}_T$. The lepton or jet isolation is selected using
a cut on $\Delta R=\sqrt{\Delta\phi^2 + \Delta\eta^2}$, where $\Delta\phi$
and $\Delta\eta$ are the difference of azimuthal angle and rapidity 
respectively, between two jets or one jet and one lepton. 
   
\begin{enumerate}
\item
Number of jets, $n_j \ge$ 2, if they satisfy the $p_T^j >$20 ,  
$|\eta_j|<$2.5 with $\Delta R(j,j)>$ 0.7. We require that 
there be at least one tagged b jet.    

\item
Leptons are selected, if $p_T^\ell >$10 and $|\eta_\ell| <$2.5
with $\Delta R(\ell,j)>$ 0.5.

\item
We require missing energy $p{\!\!\!/}_T >$ 25 .

\item
We require azimuthal angle between lepton and the direction of 
missing momentum  be 
$45^0$ $<\Delta \phi(\ell,p{\!\!\!/}_T)<$ $160^0$. 
\item
We demand $H_T < $ 120 , where $H_T = p_T^\ell + p{\!\!\!/}_T$.
\end{enumerate}

Here cuts (1-3) are event selection cuts where as the cuts (4-5) are 
for background rejection. We have imposed comparatively 
harder cuts on jets
to minimize the background from W boson production. We have noticed that
among the isolated jets, almost one is always a taggable b jet.

In Table 1, we show the response of the background processes to these cuts. 
In background processes, the source of lepton and 
invisible momentum are  W decays. 
Since, in the $t \bar t$ background process, 
particularly,
eq.\ref{ttbg}, the final state particles are expected to be more 
energetic relatively to the particles in the signal, 
the cut [5] kills the $t \bar t$ background by $\sim 30$\%
without much affecting the signal for 
$m_{\tilde t_1} \sim$100 . However, it is evident from Table 1, 
the requirement of $H_T<$ 120 
is not helpful to  
suppress the SM backgrounds from $W$ boson production with or 
without jets. The signal cross section, as for example, 
for $m_{\tilde t_1}=$120 and $m_{\tilde\chi_1^0}=$60, turns out 
to be 0.39 $\it {pb}$. Evidently, the level of signal cross section is much 
below the level of background. 
To reject backgrounds further, we exploited another kinematic variable,
the transverse mass constructed from the transverse momentum and
and missing energy ~\cite{tata} :
\be 
m_T =\sqrt{2 p_T^{\ell} p{\!\!\!/}_T 
(1 - \cos[\Delta\phi(\ell,p{\!\!\!/}_T)}]
\label{MT}
\ee 
Fortunately, in top and W backgrounds, 
the lepton and much of the invisible momentum originate from a single
W decay leading to a Jacobian peak which is absent in the signal 
process. We studied and found that the $m_T$ distribution for the signal
is highly populated towards lower values. 
Therefore, by demanding $m_T<$45, the background 
cross section is reduced severely. This is reflected in Table.1, 
in the the last row. 
As for example, the most dominant background after cuts
[1-5] is from $W+$2 jets. It is suppressed by one order of magnitude 
because of $m_T$ cuts where  as $t\bar t$ is suppressed by a factor 
of $\sim$3 with a modest loss in signal cross section. 
As we see from the last row, the combined background cross section
turns out to be 0.22 pb leading to  $\sim$ 440 events per year with the
expected lumonisity 2 fb$^{-1}$ per year.   

\begin{table}[!htb]
\begin{center}
\begin{tabular}{|c|c|c|c|c|}
\hline
Cuts& $t \bar t$ & $W+2$jets & $WW$ & $WWj$  \\
\hline
[1-3]&.82 & 11 & 1 & .13 \\
\hline
[4]&.32 & 8 & .72 & .10 \\
\hline
[5]&.25 & 7 & .64 & .09 \\
\hline
$m_T<$45 GeV &.08 & .06 & .07 & .01 \\
\hline
\end{tabular}
\vspace*{-2mm}
\end{center}
\caption{The Background cross sections (in $\it {pb}$) after implementation
of kinematical cuts described in the text. Respective branching ratio 
suppression factors are included.}
\end{table}

We have computed the signal cross section for  
some representative values of
r=2, 1.5, 1.3. In Fig.6, we present the signal cross
sections(solid lines) subject to all kinematic cuts as described above
including the cut on $m_T$. The suppression due to the BR of 
4-body decay is not included. This will be discussed below in the 
context of different models. For comparison, we also give the
stop pair production cross sections(dashed lines) in the same figure.
Recall that the hardness of the final state particles in the signal
depends on $\Delta m$. Larger values of $\Delta m$ mean,
higher value of acceptance efficiency. This explains, the rapid decrease 
of signal cross section as shown in Fig.6 for smaller value of
r, say 1.3.

To convert this signal cross section into event rates, we have to 
multiply the signal cross sections in Fig.6 by the respective branching ratio 
suppression for $\tilde t_1$ decay, i.e by $\epsilon_{br}$ =2. 
Br$(\tilde t_1 \ar b \chi_1^0 \ell \nu_\ell)$.
Br$(\tilde t_1 \ar b \chi_1^0 q \bar q')$ ($\ell=$e,$\mu$). 
In Sec.III, we have discussed the decay rates of stop in these channels
for representative values of SUSY parameters. It follows from Fig.3-5 
that the value of $\epsilon_{br}$ may vary from a few percent
to $\sim$ 20\%.
For the parameter set in Fig.3 we find that $m_{\tilde\chi_1^0} 
\approx $87. Thus for
$\MST$=135, r $\approx$1.5. For this $\MST$, $\epsilon_{br}$ 
$\simeq$ 20\%.
On the other hand from Fig.6 the corresponding signal cross-section is 
0.09 pb. Putting all these together one can expect for 2 fb$^{-1}$ 
of integrated luminosity approximately 36 signal events (S). From 
Table 1 the total number of background  events (B) is estimated to be 
440, leading to a rather modest S/$\sqrt B$ ratio.

To reject the background further,
we can impose an additional requirement in event selection by tagging
one of the jets as a b jet. We know, the major upgradations of CDF 
and D0 detectors are underway. Among the improvements 
the capability to trigger on displaced vertices from b quark decays using
a precise microvertex detector, is very encouraging. 
It may be possible, to achieve a sufficient b-tagging efficiency, like
$\sim$50\% or so~\cite{sugra}.
Therefore, exploiting
this facilities we can improve S/$\sqrt B$ by demanding at least 
one b jet
in signal events at the cost of paying a price for b-tagging
efficiency in the event number.  
This requirement leads to suppression of  the backgrounds from $W$ boson
production to an enormous amount as there is no 
b jet in the resulting final state. However, there is a chance that an ordinary
quark jet may fake as a b jet, but this probability is very very 
small~\cite{mistag}. The, requirement of at least one tagged b jet
in the signal will help to get rid of the background process, 
except for the ones from $t \bar t$, which is not at a negligible level. 
Nevertheless
we gain in  S/$\sqrt B$ by 
paying a price for b-tagging efficiency  
by a factor of 2 in the number of signal events.

Turning to the mSUGRA parameter set in Fig.4 with $\MLSP \simeq$ 54 we find that for $\MST \approx$100, r $\simeq$2. A relatively large signal cross-section
as given by Fig.6 leads 100 events in spite of the suppression factor 
$\epsilon_{br} \simeq$ 20\%, as can be read off from Fig.4 for tan$\beta$=3.
The S/$\sqrt B$ ratio thus obtained is quite encouraging even without 
b tagging. From Fig.5 it is clear that $\epsilon_{br} \simeq $ 20\% 
can be easily realized
in the low tan$\beta$ scenario. 

We now discuss the discovery limit of $\t1$ in the channel considered 
in a fairly model independent way. We have seen above that 
$\epsilon_{br} \approx $ 20\% 
can be realised in a variety of models with tan$\beta$=3-4 and this 
 will be used in our analysis as a representative value. It is now
easy to see from Fig.6 
that for $\MST$=120 one ends up with S/$\sqrt B$= 4(2)
for r $\simeq$2(1.5) and an 
integrated luminosity of 2 fb$^{-1}$. For r $\simeq$2 the search limit
can be extended to $\sim$150 for this $\epsilon_{br}$.

The prospect looks even better  for the 
high luminosity option with an integrated luminosity  $\sim$ 15 fb$^{-1}$ 
which is expected to accumulate 
after few years of running. In that option, discovery limit of 
$m_{\tilde t_1}$ might be better in the proposed channel, even for lower values of r and $\epsilon_{br}$.

Before ending this section, we note that there is 
a possibility of detecting the 4-body decay of $\tilde t_1$  through the jet 
plus missing energy channel with or without b-tagging. This final state occurs
in stop pair production when both the stops decay hadronically,
eq.~\ref{4dk}. This channel looks promising since the corresponding suppression  factor $\epsilon_{br}$ is rather mild. However in considering this channel, one has to worry about the huge QCD background. Nevertheless, it is 
worthwhile to examine the observability of $\tilde t_1$ in this channel. We
have not discussed this in our present analysis as it is beyond  
the scope of parton level calculations.  

\begin{figure}[!t]
\vspace*{-3.5cm}
\hspace*{-3.0cm}
\mbox{\psfig{file=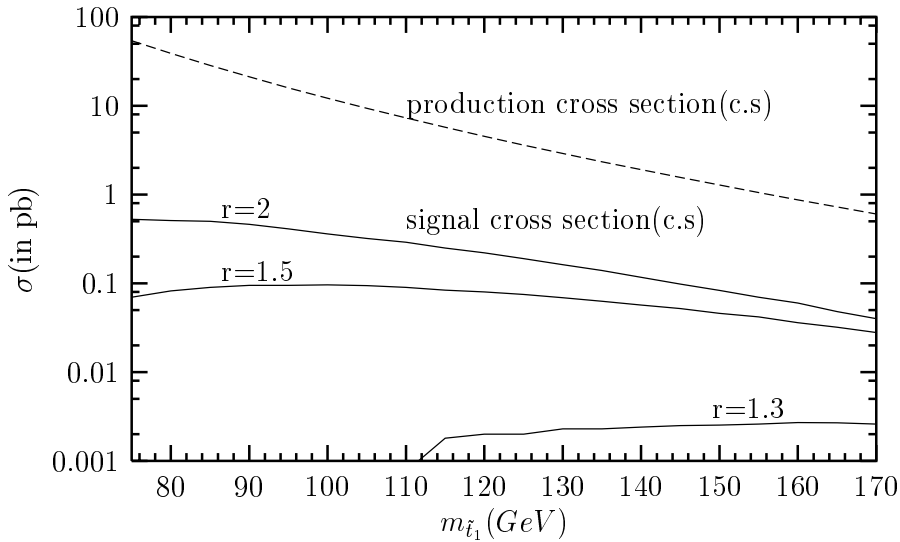,width=20cm}}
\vspace*{-16.7cm}
\caption{\small 
Cross sections(pb) for stop pair production(dashed line) and
signal cross section(solid lines) subject to all kinematical cuts, 
including cut on $m_T$, for some representative values of $r$,
where $r=\frac{m_{\tilde t_1}}{m_{\tilde\chi_1^0}}$. 
Here the branching ratio suppression due to the stop decays is not included.
}
\label{fig_mT}
\end{figure}

\section*{V.~Conclusions}

In conclusion we reiterate that the 4-body decay of $\tilde t_1$
~\cite{boehm,djouadi}
may indeed play an important role in stop searches at the upgraded Tevatron
if it happens to be the NLSP. Especially for low values of 
$\tan\beta$(=3 - 4)
this may be the main discovery channel.

We have delineated the mSUGRA parameter space where the $\tilde t_1$ 
is the NLSP. 
For a given $m_0$, $\mhf$, tan$\beta$, sign($\mu$), a range of values of the trilinear coupling $A_0$ yields a stop NLSP.

We have studied the BR of the four body decay of $\tilde t_1$ in both 
mSUGRA and MSSM models. We found it to be numerically significant and 
sometimes even the dominant decay mode in a large region of parameter 
space with low tan$\beta$
(3 $\lsim$ tan$\beta$ $\lsim$ 5). Here the much studied loop induced 2-body 
decay of $\tilde t_1$ is suppressed (Fig.3-5). The 4-body leptonic BR of 
$\tilde t_1$ which is an essential ingredient of stop search in the 
channel proposed by us is also found to be appreciable.

In view of the 4-body decay the current limits on $\MST$ from Tevatron 
RUN-I based on the assumption that the loop induced decay of $\tilde t_1$ 
occurs with 100\% BR,
needs revision and becomes somewhat model dependent. It follows from 
Fig.4 that the current limit $\MST \gsim$102 for $\MLSP$=50 will
be significantly relaxed for 
$\tan\beta \approx$ 3, since the corresponding BR of the loop induced 
decay is indeed small. For 3$\lsim \tan\beta \lsim$5, the 4-body decay is 
likely to have nontrivial impact on stop searches at the upgraded 
Tevatron. For 
$\tan\beta \gsim$ 5 the loop decay practically overwhelms the 4-body 
decay. For such values of $\tan\beta$ the curent limits from 
Tevatron RUN-I are valid 
and strong upper limits on $|A_0|$ can be placed in mSUGRA models 
for a given set of SUSY 
parameters as is clearly seen from Fig.1 and 2. These limits are more 
restrictive than the ones obtained from the CCB condition or from the 
requirement that $\LSP$ be the LSP. Thus while the constraint on 
the $m_0$-$\mhf$ plane from squark-gluino searches ~\cite{xsusy} 
is fairly independent of $A_0$, this parameter can be strongly constrained 
from the negative result of stop search for each allowed
$m_0$-$\mhf$ pair.

We studied the viability of discovering the stop in the leptons+jets+$\ET$
channel which arise when one of the stops decays leptonically and the 
other decays hadronically via the 4-body mode. We have listed the 
Standard Model 
backgrounds and the kinematical cuts ( see the Table ) to suppress them.
This discovery limits sensitively depends on 
$\epsilon_{br}$ 
which are model 
dependent in general and on $r$. Our studies of the BR in a variety 
of models in 
Sec.III indicate that 
$\epsilon_{br} \approx $20\% is a fairly representative choice. Armed with 
this information we have estimated the discovery limit to be $\MST$=120(150)
for r=1.5(2) for an integrated luminosity 2 fb$^{-1}$ without requiring 
b-tagging. We have discussed qualitatively how the search prospect improves if
b-tagging and higher accumulated luminosity after the proposed luminosity 
upgrade are available. The possibility of stop search in the jets+$\ET$ 
channel which has a less severe branching ratio suppression($\epsilon_{br}$) 
is also discussed.

\vspace*{5mm}
{\bf Acknowledgement}:

AD acknowledges the grant of a DST, India Research Project (No.\ SP/S2/k01/97). SPD's work was supported by a fellowship from Council of Scientific 
and Industrial Research(CSIR), India. MG is grateful to A. Djouadi
for drawing attention in this topic and he is also 
thankful to the theory group of CERN where part of this project was 
carried out.

\end{document}